\documentclass{article}

\usepackage{microtype}
\usepackage{graphicx}
\usepackage{subcaption}
\usepackage{booktabs} % for professional tables

\usepackage{hyperref}

\usepackage[preprint]{icml2026}

\usepackage{amsmath}
\usepackage{amssymb}
\usepackage{mathtools}
\usepackage{amsthm}
\usepackage{tikz}
\usepackage{listings}
\usetikzlibrary{shapes,arrows,positioning,fit,backgrounds}

\lstdefinelanguage{Coq}{
  keywords={Example, let, in, Proof, Admitted, Defined, Qed, Definition, Theorem, Lemma,
            Fixpoint, CoFixpoint, match, with, end, forall, fun, exists, if, then, else,
            return, as, struct, Record, Inductive, CoInductive, Type, Prop, Set,
            Module, Section, End, Import, Export, Require, From, Variable, Hypothesis,
            Parameter, Axiom, Notation, Check, Compute, Print, Eval},
  sensitive=true,
  morecomment=[s]{(*}{*)},
  morestring=[b]",
}

\usepackage[capitalize,noabbrev]{cleveref}

\theoremstyle{plain}

\theoremstyle{definition}

\theoremstyle{remark}

\usepackage[textsize=tiny]{todonotes}

\icmltitlerunning{Submission and Formatting Instructions for ICML 2026}

\begin{document}

\twocolumn[
  \icmltitle{Neuro-Symbolic Generation and Validation of Memory-Aware Formal Function Specifications}



  \icmlsetsymbol{equal}{*}

% Liao Zhang, Tong Chen, Xiwei Wu, Qi Liu, Xiyu Zhai, Xinqi Wang, Qinxiang Cao 

  \begin{icmlauthorlist}
    \icmlauthor{Liao Zhang}{sj-cs}
    \icmlauthor{Tong Chen}{shii}
    \icmlauthor{Xiwei Wu}{sj-cs}
    \icmlauthor{Qi Liu}{shii}
    \icmlauthor{Xiyu Zhai}{uw}
    \icmlauthor{Xinqi Wang}{uw}
    \icmlauthor{Qinxiang Cao}{sj-ai}
  \end{icmlauthorlist}

  \icmlaffiliation{sj-cs}{School of Computer Science, Shanghai Jiao Tong University}
  \icmlaffiliation{sj-ai}{School of Artificial Intelligence, Shanghai Jiao Tong University}
  \icmlaffiliation{shii}{Shanghai Innovation Institute}
  \icmlaffiliation{uw}{School of Computer Science and Engineering, University of Washington}

  \icmlcorrespondingauthor{Liao Zhang}{zhangliao714@gmail.com}

  \icmlkeywords{Formal Verification, Separation Logic, Neuro-Symbolic Systems, Program Specification, Memory Safety, Large Language Models}

  \vskip 0.3in
]



\printAffiliationsAndNotice{}  % no special notice (required even if empty)

\begin{abstract}
  Formal verification of memory-manipulating programs critically depends on precise function specifications that capture memory states written by experts.
  This requirement has become a major bottleneck as large language models (LLMs) increasingly generate low-level systems code whose correctness cannot be assumed.
  To enable scalable formal verification, we focus exclusively on function specification generation, deliberately avoiding the synthesis of complex loop invariants that are central to traditional verification pipelines.
  We propose a neuro-symbolic framework for automatically generating memory-aware formal function specifications for C programs from natural language problem descriptions and function signatures.
  The pipeline first produces candidate specifications via in-context learning, and then iteratively refines them using compiler diagnostics from symbolic provers and the verification toolchain.
  In particular, we validate candidate specifications by constructing a proof for the negation of the specification with concrete examples, enabling machine-checked rejection of plausible-but-incorrect specifications.
  To support systematic evaluation, we introduce LeetCode-C-Spec, a new benchmark of 200 C programming problems for generating memory-aware formal function specifications.
  Experiments show that iterative refinement substantially improves syntactic validity, while symbolic prover-based refutation significantly enhances correctness assessment by filtering false positives that LLM-only judges frequently accept.
  Our results demonstrate that combining neural generation with symbolic feedback provides an effective approach to formal specification synthesis for memory-safe systems software.
\end{abstract}

\section{Introduction}

The advent of large language models (LLMs) has fundamentally transformed software development, enabling automated code generation at unprecedented scale.
Systems like Codex , GPT-4 , Code Llama , and StarCoder  can synthesize non-trivial programs from natural language, and are already integrated into widely-used development tools including GitHub Copilot , Cursor , and Claude Code .
Due to the probabilistic nature of neural code generation, modern software pipelines can now produce code faster than software engineers can reliably trust it.

This gap is especially acute for systems code: LLM-generated programs may look correct but still contain subtle errors, including memory safety violations and brittle edge-case logic .
For programs used in security- and safety-critical settings, a single specification mismatch can manifest as exploitable vulnerabilities.
Hence, we need mechanisms that provide machine-checkable guarantees rather than probabilistic confidence.

\begin{figure}[t]
  \centering
\begin{lstlisting}[language=C, basicstyle=\scriptsize\ttfamily, breaklines=true]
/*@ Extern Coq (rev: list Z -> list Z) */

struct list *reverse(struct list *p)
/*@ With (l: list Z)
    Require sll(p, l)
    Ensure sll(__return, rev(l)) */
{
    struct list * w = (void *) 0, * v = p;
    /*@ Inv Assert
    exists p_v l1 l2, l == app(rev(l1), l2) &&
    sll(w, l1) * sll(v, l2) * data_at(&p, p_v) */
    while (v) {
      struct list * t = v -> next;
      v -> next = w;
      w = v;
      v = t;
    }
    return w;
}
\end{lstlisting}
\caption{QCP formal verification for singly-linked list reversal with separation logic. The specification includes precondition \texttt{Require sll(p, l)} and postcondition \texttt{Ensure sll(\_\_return, rev(l))}. The loop invariant maintains that the concatenation of reversed prefix and remaining suffix equals the original list. The separation operator ($*$) ensures disjoint heap regions for memory safety.}
\label{fig:qcp-sll-verification}
\end{figure}

Formal verification is a promising path to provide guarantees for program correctness, but it hinges on a missing ingredient: \emph{formal specifications}. A specification precisely states what a program should do, typically as \emph{function specifications} (preconditions and postconditions ) together with \emph{loop invariants} for programs with loops or recursion . For memory-manipulating C code, specifications must also capture memory safety, often using separation logic .

In practice, writing these formal specifications is a major bottleneck that prevents verification from keeping pace with the rapid adoption of LLM-generated code. Traditional specification writing requires deep expertise in both the problem domain and formal logic systems. This creates a significant barrier: even experienced programmers may struggle to articulate precise memory-safety requirements in separation logic, while formal verification experts may lack domain knowledge about the intended program behavior. This expertise gap becomes especially acute when dealing with LLM-generated code, where the volume of code production vastly exceeds the capacity for manual specification writing.
To illustrate these concepts, consider verifying a function that reverses a singly-linked list (see Figure~\ref{fig:qcp-sll-verification}). The precondition must state that the input pointer points to a valid linked list: in the example, this is expressed as \texttt{Require sll(p, l)}, where \texttt{sll(p, l)} asserts that pointer \texttt{p} points to a singly-linked list with elements \texttt{l}. The postcondition must guarantee that the output forms a reversed list: here, \texttt{Ensure sll(\_\_return, rev(l))} ensures the return value is a valid list containing the reversed sequence.
The loop invariant, beginning with \texttt{Inv Assert}, must express that the partially reversed prefix \texttt{l1} and remaining suffix \texttt{l2} together reconstruct the original list \texttt{l}.
Finally, memory safety is enforced through the separation logic predicate \texttt{sll}, which ensures that each list node occupies a distinct, non-overlapping memory region.

Since writing formal specifications is a challenging task, it is natural to leverage large language models (LLMs) to assist with specification generation.
Recent work has begun exploring LLM-based approaches to formal verification and specification synthesis.
However, these approaches face a fundamental challenge: the synthesis of loop invariants is a well-studied bottleneck in formal verification, and is in general undecidable or highly complex to automate.

Our goal is to make memory-aware specification writing for C programs scalable.
Given a natural language problem description and a function signature, we aim to automatically generate a separation-logic function specification that is not only syntactically valid, but also robust under adversarial validation.
Our key insight is that while proving the correctness of a candidate specification is difficult, refuting an incorrect one is often feasible by constructing concrete counterexamples via test execution combined with automated theorem proving.
Moreover, function specifications describe input-output relationships at the function boundary, whereas loop invariants must hold inductively across all iterations and typically require deeper semantic reasoning.
This distinction makes automated loop invariant synthesis significantly harder than function specification generation, motivating our focus on function-level specifications.

In contrast to previous work, we focus solely on automatically generating memory-aware function specifications for C programs, deliberately leaving loop invariant generation out of scope. This design choice enables (1) validation through test case execution without full mechanized proofs, (2) layered verification where validated specifications serve as foundations for deriving loop invariants, and (3) reinforcement learning based improvement via test-based reward signals.

We target C programs and the QCP verification toolchain, which supports separation-logic specifications and exports complex proof obligations to Coq.
We use LeetCode problems as they provide standardized data structures, input-output formats, and explicit test cases, with substantially higher algorithmic complexity than existing C verification benchmarks.

\textbf{Our Contributions.}
  \begin{itemize}
  \item We propose an agentic workflow for generating memory-aware C function specifications, combining iterative refinement guided by compiler feedback to ensure syntactic correctness with counterexample-guided refutation via automated theorem proving for correctness assessment.
  \item We construct LeetCode-C-Spec-200, a benchmark of 200 LeetCode problems for formal C specification generation, featuring more complex algorithms than prior C verification benchmarks.
  \item We empirically demonstrate that the proposed agentic workflow significantly improves the syntactic validity of LLM-generated function specifications. Furthermore, our neuro-symbolic validation substantially enhances correctness assessment, with Coq-based refutation effectively filtering false positives that are sometimes accepted by LLM-only judges.
\end{itemize}

\section{Preliminaries}

In this section, we provide background on separation logic and annotation-based program verification, which form the foundation of our approach to generating memory-aware formal specifications.

\subsection{Separation Logic}

Separation logic  is a formal system for reasoning about programs that manipulate pointers and dynamic memory. Its key innovation is the \emph{separating conjunction} ($*$), which asserts that $P * Q$ holds when memory can be split into two non-overlapping regions satisfying $P$ and $Q$. Crucially, $\texttt{store}(p, v) * \texttt{store}(q, u)$ implies $p \neq q$, enabling local reasoning without aliasing concerns.

The singly linked list predicate $\texttt{sll}(p, l)$ relates pointer $p$ to logical list $l$:
\begin{multline*}
\texttt{sll}(p, l) :=
((l = \texttt{nil} \land p = \texttt{null}) \land \texttt{emp})
\;\;\lor \\
(\exists v, l', p'.\;
(l = v :: l') \land
\texttt{store}(\texttt{addr}(p.data), v) * \\
 \texttt{store}(\texttt{addr}(p.next), p') * \texttt{sll}(p', l')).
\end{multline*}
The base case handles empty lists, while the recursive case uses $*$ to enforce disjoint memory for each node.

\subsection{Annotation-Based Program Verification}

Annotation-based verification  augments source code with formal specifications as annotations. A verifier generates \emph{verification conditions} (VCs)—logical formulas whose validity implies correctness—which can be discharged by SMT solvers or interactive theorem provers like Coq.

We use QCP , a state-of-the-art C verifier supporting separation logic predicates and exporting complex VCs to Coq for manual or semi-automated proving.

QCP predicates are defined in Coq. Predicates not in QCP's standard library must be imported via \texttt{Extern Coq} declarations (e.g., \texttt{rev} in Figure~\ref{fig:qcp-sll-verification}).

\section{Specification Generation}

\begin{figure*}[t]
\centering
\includegraphics[width=0.95\textwidth]{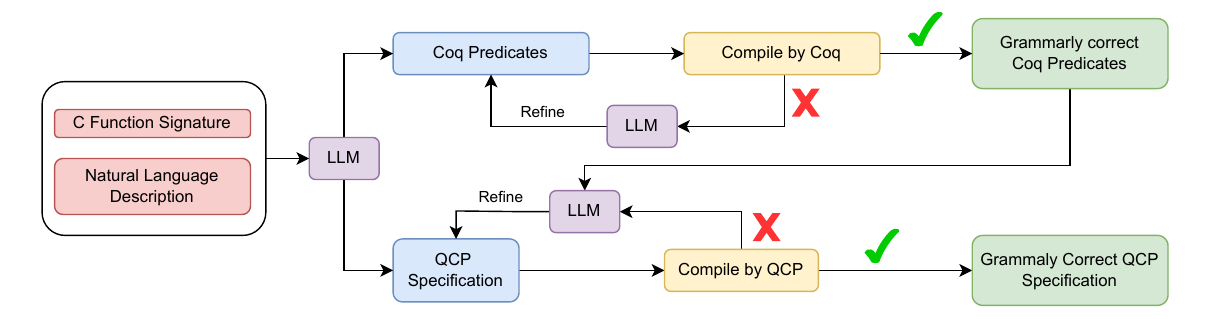}
\caption{System architecture for LLM-based generation and refinement of memory-aware formal specifications. The pipeline consists of two main phases: (1) \textbf{Specification Generation} using LLM with in-context learning, and (2) \textbf{Iterative Refinement} guided by compiler feedback, symbolic execution, and counterexample-guided refutation to ensure both syntactic correctness.}
\label{fig:system-architecture}
\end{figure*}

Figure~\ref{fig:system-architecture} presents the overall architecture of our LLM-based formal specification generation and validation system. The system takes as input a natural language problem description and a C function signature, and produces formal specifications and additionally necessary Coq predicates. The pipeline consists of two main phases: initial specification generation and iterative refinement guided by compiler feedback and counterexample-guided refutation.

\textbf{Phase 1: Initial Specification Generation.}
The generation phase employs prompt engineering with carefully curated examples to guide the LLM in producing syntactically and semantically correct QCP specifications.

Our prompts incorporate three key components.
First, we provide natural-language descriptions of QCP’s predefined predicates, including their types, parameters, and intended usage contexts.
For example, we describe \texttt{tree\_insert(v: Z, t: tree): tree} as ``inserts a value \texttt{v} into a binary search tree \texttt{t} and returns the updated tree."
These predicates are manually selected from QCP's standard library based on their relevance to LeetCode problems, with a focus on commonly used data structures such as arrays, linked lists, and binary trees.
We do not describe all predicates in QCP's standard library, as many are irrelevant to formal function specification generation (e.g., auxiliary predicates such as \texttt{increasing\_aux}).

Second, we include few-shot examples demonstrating correct specification patterns for common data structures (arrays, linked lists, trees) and operations. These examples are drawn from three sources: (1) QCP's standard test suite, which includes fundamental operations such as list reversal and binary search tree insertion; (2) several synthetic problems inspired by algorithmic patterns common in competitive programming but not directly from LeetCode; and (3) one representative LeetCode problem that is excluded from our evaluation benchmark to avoid data leakage. The complete prompt template is provided in Appendix.
% ~\ref{app:refutation}.

Third, we incorporate common error patterns that summarize frequent syntactic mistakes made by LLMs when generating separation logic annotations. These error patterns—derived from empirical analysis of failed generation attempts—include issues such as incorrect predicate instantiation, malformed separation conjunctions, type mismatches in Coq predicates, and improper use of quantifiers. By providing predicate documentation, positive examples, and negative patterns, we significantly reduce the rate of trivial syntactic errors in initial generations.

% We additionally introduce a dedicated step for importing the external QCP predicates required by each task. This reflects our design principle for agentic workflows: decompose a complex end-to-end objective into simpler sub-tasks that are easier for an LLM to solve with a shorter context window, thereby improving overall reliability.

\textbf{Phase 2: Iterative Refinement with Symbolic Feedback.}
When the generated specification fails to compile or is found to be semantically incorrect, we apply an iterative refinement loop. This loop is guided by tool feedback from both Coq and QCP (addressing syntax and type errors) as well as formal refutation (addressing semantic errors).
For user-defined Coq predicates, we leverage Coq compiler diagnostics to localize syntax errors, type mismatches, and unresolved identifiers; we pair these diagnostics with curated repair examples in the prompt to steer the model toward minimal fixes.
For QCP annotations, we collect diagnostics from the QCP checker and categorize failures into two primary cases.
(1) \emph{Annotation syntax errors}: We utilize a taxonomy of common QCP mistakes and corresponding repair patterns to guide the model. Specifically, we target two frequent error classes: (a) language confusion between QCP and Coq syntax (e.g., using \texttt{=} instead of QCP's \texttt{==}); and (b) QCP-specific annotation mistakes, such as malformed \texttt{Require}/\texttt{Ensure} clauses or incorrect predicate instantiation. The full error taxonomy and repair patterns are detailed in Appendix.
% ~\ref{app:syntax-errors}.
(2) \emph{Predicate import errors}: To resolve these, we first strip existing import annotations from the C file, then employ a dedicated prompt to regenerate the required \texttt{Extern Coq} declarations from the synthesized Coq definitions. These are then reinserted into the original C file alongside the QCP specification.
Errors outside these categories are currently treated as unrecoverable and lead to generation failure.

\section{Specification Validation}

Validating LLM-generated specifications presents a fundamental challenge: determining whether a specification correctly captures the intended program behavior.
We address this challenge through a multi-method validation framework that combines formal proof-based techniques with LLM-based semantic reasoning.

While LLM judges can assess high-level semantic alignment between specifications and natural language requirements, they struggle to reliably validate the intricate logical relationships encoded in separation logic predicates. For instance, an LLM may fail to detect that a specification incorrectly allows aliasing between supposedly disjoint heap regions, or that a postcondition is too weak to establish necessary memory-safety guarantees. Conversely, formal methods provide rigorous guarantees but require concrete test cases to instantiate symbolic specifications.

Our key insight is that input-output test cases—ubiquitous in competitive programming and software testing—can bridge this gap by providing concrete program behaviors against which specifications can be formally checked. Each test case grounds abstract separation logic assertions in concrete heap configurations, enabling automated theorem provers to either validate the specification or discover counterexamples that definitively refute it.

\subsection{Counterexample-Guided Refutation}

A fundamental asymmetry exists between proving and disproving specifications: establishing correctness requires reasoning over all possible inputs, whereas refuting it requires only a single counterexample. We exploit this asymmetry through a refutation-based validation strategy. While passing validation on many test cases provides no guarantee of correctness, finding a single input-output pair that violates the postcondition definitively proves the specification incorrect.

Formally, let $\mathtt{post}$ denote the postcondition of a candidate specification. The specification is incorrect if there exists an input-output pair $(in, out)$ that violates the postcondition:
\begin{align}
\exists in, out \cdot \neg\mathtt{post}(in, out)
\end{align}
For leetcode problem, every problem several examples (mostly 2 or 3) are provided. We can direcly use those examples in generating $\neg\mathtt{post}(in, out)$.

\begin{figure}[t]
  \centering
  \includegraphics[width=\columnwidth]{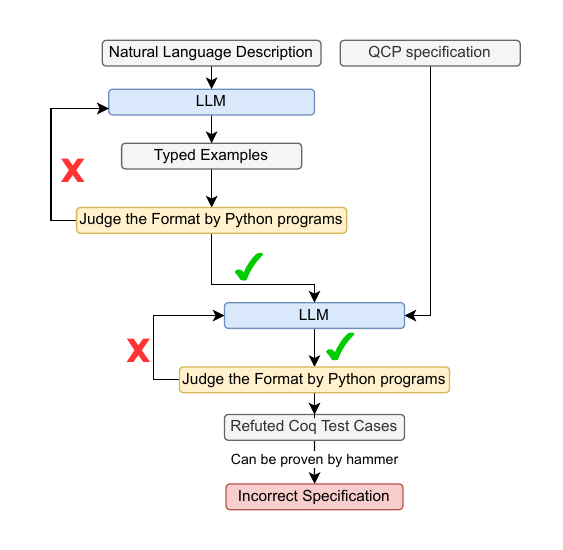}
  \caption{Counterexample-guided refutation workflow. Given a candidate specification, we negate the postcondition and attempt to find a test case satisfying the precondition but violating the postcondition. If such a counterexample exists, symbolic execution in Coq can concretely demonstrate the specification's incorrectness by exhibiting the violating heap state.}
  \label{fig:refute}
\end{figure}

We implement this refutation strategy by combining LLM-based test case generation with automated theorem proving in Coq, as illustrated in Figure~\ref{fig:refute}. The workflow proceeds in two stages:

\paragraph{Negation and Test Case Generation.} Given a specification with postcondition $\mathtt{post}$, we construct the negated verification goal $\neg\mathtt{post}$ and generate concrete test cases—specific heap configurations and input values—designed to expose potential specification errors.

Since input examples can be complex, directly prompting LLMs to convert natural language descriptions into correct Coq representations may introduce errors. We employ a three-stage pipeline to improve reliability. 
First, we prompt an LLM to type the natural language examples, producing structured typed examples with explicit type annotations. The typed example 
Second, we apply deterministic Python programs to translate these typed examples into canonical Coq format, ensuring syntactic correctness. Third, we provide the LLM with both the typed Coq examples and the original natural language description to generate refutation test cases.

\begin{figure}[t]
  \centering
  \begin{minipage}{\columnwidth}
  \begin{lstlisting}[language=Coq, basicstyle=\scriptsize\ttfamily, breaklines=true]
Example example2:
  let tr := empty in
  let val := 4 in
  let result := tree_insert (val + 1) tr in
  let expected := (make_tree empty 4 empty) in
  result <> expected.
Proof.
Admitted.
  \end{lstlisting}
  \end{minipage}
  \caption{Example format of a Coq refutation test case (tree insertion). The test binds concrete inputs, computes \texttt{result}, and asserts it differs from the expected output.}
  \label{fig:refute-testcase-format}
\end{figure}

We use a carefully designed prompt to convert QCP specifications into refutation test cases in Coq.
The prompt provides several examples, each following the same mechanized structure to facilitate the conversion.
Each example begins with \texttt{let arg := ...} to bind each input variable to concrete values. After introducing the input variables, it uses \texttt{let result := ...} to compute the output obtained from the specification. Finally, we assert that this result is not equal to the expected output value from the LeetCode problem description.
% An example of the refutation example is shown in the Appendix~\ref{app:refut-exg}.

Figure~\ref{fig:refute-testcase-format} illustrates the standard format of a refutation example.
Here, \texttt{tr} and \texttt{val} denote the input tree and the value to be inserted, respectively.
The variable \texttt{result} represents the tree produced by the insertion function, while \texttt{expected} denotes the tree described by the natural language specification.
The goal asserts that \texttt{result} is not equal to \texttt{expected}, thereby refuting an incorrect candidate specification.

To ensure correctness, we apply automated sanity checks before validation: (i) for each input variable \texttt{x} appearing in the natural language test case, we verify a corresponding \texttt{let x := ...} binding exists in the generated Coq file; and (ii) we reject any generated code containing \texttt{admit}, \texttt{Parameter}, or \texttt{Axiom} declarations that would trivialize the proof.

\paragraph{Automated Counterexample Validation.} For each candidate test case, we instantiate the negated specification and attempt to discharge the proof goal using Coq's \texttt{hammer} tactic . Hammer bridges Coq with external automated theorem provers (ATPs) including Z3 , E prover , CVC4 , and Vampire . It translates Coq goals into first-order logic, invokes these ATPs to find proofs, and reconstructs successful proofs in Coq's native format. If hammer discharges the negated goal for any test case, the specification is formally refuted—the test case serves as a machine-checked counterexample. While hammer cannot prove all theorems, it has demonstrated strong performance on verification-related goals in large-scale projects .

% This refutation-based approach provides several advantages: (1) it can definitively reject incorrect specifications without requiring full verification, (2) it generates interpretable counterexamples that explain \emph{why} a specification is wrong, facilitating refinement, and (3) it complements proof-based validation by catching semantic errors that may not manifest as compilation failures.

\section{Dataset Construction}

A major bottleneck in evaluating LLM-based formal specification generation is the scarcity of suitable benchmarks. Existing C programming datasets often lack formal function signatures, focus primarily on implementation correctness, or are limited in scale and algorithmic diversity. To address these limitations, we introduce \textsc{LeetCode-C-Spec}, a comprehensive benchmark designed specifically for memory-aware formal specification generation.

We first scraped 2,766 C-language problems from LeetCode, capturing natural language descriptions, input/output examples, and function signatures. We then manually sanitized the problem descriptions and signatures to ensure syntactic consistency. Unlike existing datasets that prioritize solution code, our benchmark focuses on function signatures as the primary target for specification generation, which necessitates reasoning about parameter types, return values, and implicit memory layout assumptions.

From the initial collection, we identified 439 problems (with IDs below 837) whose data structures are compatible with QCP's built-in separation logic predicates.
To standardize the problems, we manually removed constraints from the problem descriptions that are irrelevant to functional (input-output) correctness.
These constraints (e.g., \texttt{0 <= m, n <= 200}) are introduced primarily to bound execution time or memory usage on the LeetCode platform, and do not affect the intended semantic behavior of correct solutions.

We further restrict our dataset to problems with a single return value, as supporting multiple output values would require additional specification machinery that is orthogonal to the focus of this work. This restriction still preserves substantial complexity: many problems involve intricate memory operations, such as in-place array manipulation, tree restructuring, and linked list transformations.

Despite this restriction, the selected problems involve non-trivial algorithms and complex data structures, and are substantially more challenging than benchmarks used in prior work on C program verification.
Prior benchmarks typically focus on simple algorithms like list reversal, array copying, or basic tree traversals. In contrast, our benchmark includes algorithmic challenges such as dynamic programming over arrays, graph algorithms on tree structures, and complex string manipulations requiring precise reasoning about memory boundaries and pointer arithmetic.

From this candidate pool, we randomly sampled 200 problems to form our core evaluation benchmark, \textsc{LeetCode-C-Spec-200}. This benchmark spans five data structure categories: integers, strings, one-dimensional arrays, linked lists, and binary trees. We manually verified each entry in the benchmark to ensure accuracy.

To bridge the gap between LeetCode's native C types and formal logic, we developed systematic transformations for QCP compatibility. For example, the LeetCode \texttt{TreeNode} structure is mapped to QCP's \texttt{tree} predicate, which formalizes both tree topology and node content using separation logic. For boolean types, we employ a preprocessor-based macro that aliases \texttt{bool} to integers, preserving the original LeetCode signatures while enabling formal verification. Detailed transformation rules are provided in Appendix.
% ~\ref{app:bool-transform}.

We will release \textsc{LeetCode-C-Spec-200} to the research community to support further work in automated formal verification. Table~\ref{tab:dataset-stats} summarizes the distribution of the benchmark; it reflects a realistic difficulty range, with medium problems accounting for 48.5\%, followed by easy (31.0\%) and hard (20.5\%) problems.
We classify a problem as involving a specific data structure if that type appears in any input or output argument. 
Consequently, the majority of problems (73.5\%) involve integers or numbers, underscoring the demand for arithmetic and logical reasoning. Arrays are present in 37.5\% of problems, strings in 26.5\%, binary trees in 20.5\%, and linked lists in 7.0\%. These categories are not mutually exclusive, as many problems require reasoning about multiple data structures—such as manipulating an array while returning an integer result—leading to a total percentage exceeding 100\%.

\begin{table}[t]
\centering
\caption{Distribution of the \textsc{LeetCode-C-Spec-200} benchmark across difficulty levels (as assigned by LeetCode) and data structures. Note that a single problem may involve multiple data structures.}
\label{tab:dataset-stats}
\begin{tabular}{lrr}
\toprule
Category & Count & Percentage (\%) \\
\midrule
\textbf{Difficulty} & & \\
Easy & 62 & 31.0 \\
Medium & 97 & 48.5 \\
Hard & 41 & 20.5 \\
\midrule
\textbf{Data Structure} & & \\
Integer/Number & 147 & 73.5 \\
Array & 75 & 37.5 \\
String & 53 & 26.5 \\
Tree & 41 & 20.5 \\
Linked List & 14 & 7.0 \\
\bottomrule
\end{tabular}
\end{table}

\nocite{langley00}

\bibliography{example_paper}
\bibliographystyle{icml2026}

%%%%%%%%%%%%%%%%%%%%%%%%%%%%%%%%%%%%%%%%%%%%%%%%%%%%%%%%%%%%%%%%%%%%%%%%%%%%%%%
%%%%%%%%%%%%%%%%%%%%%%%%%%%%%%%%%%%%%%%%%%%%%%%%%%%%%%%%%%%%%%%%%%%%%%%%%%%%%%%
% APPENDIX
%%%%%%%%%%%%%%%%%%%%%%%%%%%%%%%%%%%%%%%%%%%%%%%%%%%%%%%%%%%%%%%%%%%%%%%%%%%%%%%
%%%%%%%%%%%%%%%%%%%%%%%%%%%%%%%%%%%%%%%%%%%%%%%%%%%%%%%%%%%%%%%%%%%%%%%%%%%%%%%
\newpage
\appendix
\onecolumn

\end{document}